**Complicated relationships between tissue T2 relaxation time and _in vivo_ tissue diffusion measures, depending on the ranges of T2 value.**

Running title: Complicated relationships between T2 and tissue diffusion measures.

Yì Xiáng J. Wáng

Department of Imaging and Interventional Radiology, Faculty of Medicine, The Chinese University of Hong Kong, Shatin, New Territories, Hong Kong SAR, China.

E-mail: yixiang_wang@cuhk.edu.hk

## Abstract

Apparent diffusion coefficient (ADC) is a measure of the magnitude of diffusion of water molecules within tissues. We argue that ADC value contains information of both diffusion and T2 relaxation. In this letter, we list literature evidence to support our argument. Firstly, we list uterine myometrium tumors as examples. Myometrium has a T2 relaxation time of 79 ms at 3T. Literature shows, when the myometrium tumors are T2 weighted signal intensity hypertensive (relative to myometrium), these tumors likely show diffusion restriction. This is similar to that, while the spleen is T2 weighted signal intensity hypertensive relative to the liver, the spleen demonstrates diffusion restriction relative to the liver. On the other hand, when the myometrium tumors are T2 weighted signal intensity hypotensive (relative to myometrium), these tumors likely do not show diffusion restriction. However, when the myometrium tumors are very highly hypertensive such as the cases of leiomyoma cystic degeneration and myxoid degeneration, the relationship between T2 weighted signal intensity and diffusion is similar to that of a normal gallbladder, i.e., T2 weighted signal highly hypertensive without diffusion restriction.

For most of the tumors, much longer T2 may suggest a tumor is waterier. On $b$=0 images, the signal of tumor (relative to other tissues) can be dominated by T2 effect. On a high $b$-value images such as $b$=1000 images, it is possible that tumor regions are dominated by noises. In the end, the ADCs, which are determined by the slope between the signal at b=0 and the signal at $b$=1000, are also dominated by the T2 effect, with longer T2 being associated with higher ADC measure. In fact, since there is no diffusion gradient with $b$=0 images, and there are only noises on $b$=1000 (or $b$=800), ADC map is devoid of diffusion information.

Apparent diffusion coefficient (ADC) is a measure of the magnitude of diffusion of water molecules within tissues. ADC is considered to reflect tissue diffusion. The contribution of tissue T2 relaxation time to diffusion weighted image has been well noted. For example, 'T2 shine-through' refers to high signal on diffusion weighted (DW) images that is not due to restricted diffusion, but rather to high T2 signal which 'shines through' to the DW image. It is commonly considered that this T2 effect has been 'normalised out' (i.e., removed) during ADC calculation. In the case of gallbladder, despite apparent 'T2 shine-through' effect on DW image, its ADC is high. However, this assumption that T2 effect has been removed during ADC calculation may be invalid in many scenarios [1,2]. It is more likely that the ADC values contain information of both diffusion and T2 relaxation.

We have recently discussed that, considering the liver as the reference, the spleen's IVIM-*PF*, IVMI-*D_{slow}*, and ADC are substantially underestimated by DW imaging [2]. However, from anatomical and physiological viewpoints, it is more likely that the spleen has a higher diffusion than the liver [2, 3]. The spleen and the liver have similar extent of blood perfusion, and spleen may be waterier than the liver. The spleen is an organ with the function of storing blood. We proposed that, compared with the liver, longer spleen T2 relaxation time (79/61 ms for spleen and 46/34 ms for liver at 1.5T/3.0 T, respectively) contributes to the measured lower diffusion values for spleen [2, 3]. Hereby, we list more literature evidence to support our argument.

Firstly, we use uterine myometrium tumors as examples. Myometrium has a T2 relaxation time of 79 ms at 3T (117 ms at 1.5T). We searched literature which summarized the relationship between myometrium tumours' T2 weighted image signal and diffusion restriction on ADC map. Three review articles provided such information [4, 5, 6]. We grouped together their discussions and present them in table-1.

| Tissue type (scenario [number]) | Diffusion on ADC map | | | Signal on T2w Images |
|---|---|---|---|---|
| | Restricted | Unknown | Not Restricted | |
| DeMulder *et al*. LM cystic degenerated [1] | | | Yes | H hyperintense |
| DeMulder *et al*. LM Myxoid degenerated [2] | | | Yes | H hyperintense |
| Barral et al Cellular LM [3] | Yes | | | Hyperintense |
| DeMulder *et al*. Cellular LM [4] | Yes | | | Hyperintense |
| Bura *et al*. Highly cellular LM [5] | Yes | | | Hyperintense |
| Bura *et al*. leiomyosarcoma [6] | Yes | | | Hyperintense |
| Barral *et al*. Leiomyosarcoma [7] | Yes | | | Hyperintense |
| Barral *et al*. Common LM [8] | | | Yes | Hypointense |
| Bura *et al*. ordinary LM [9] | | | Yes | Hypointense |
| DeMulder *et al*. LM Hyaline degenerated [10] | | | Yes | Hypointense |
| DeMulder *et al*. LM Carneous degenerated [11] | | | Yes | Hypointense |
| DeMulder *et al*. LM- Calcific degenerated [12] | | | Yes | Hypointense |
| DeMulder *et al*. Lipoleiomyoma [13] | | Yes | | Heterogeneous |
| DeMulder *et al*. STUMP [14] | | Yes | | Heterogeneous |
| Barral *et al*. Degenerated LM [15] | | | Yes | Heterogeneous |

**Table-1. Relationship between myometrium tumor T2 weighted signal intensity and tumor diffusion restriction.** Summarized from table-2 by DeMulder *et al*. [4], table-1 by Bura *et al*. [5], and table-1 by Barral *et al*. [6]. LM: leiomyoma; STUMP: smooth muscle tumors of uncertain malignant potential. H hyperintense: highly hyperintense.

Table-1 shows, when the uterine myometrium tumors are T2 weighted signal intensity hypertensive (relative to myometrium), these tumors likely show diffusion restriction (scenarios [3-7]). This is similar to that, while the spleen is signal intensity hypertensive relative to the liver, spleen demonstrates diffusion restriction relative to the liver. On the other hand, when the tumors are T2 weighted signal intensity hypotensive (relative to uterine myometrium), these tumors likely show diffusion restriction (scenarios [8-12]). When the tumors have heterogeneous T2 weighted signal, they were noted as without diffusion restriction or diffusion restriction unknown. However, when the lesions are *very* highly hyperintense such as the cases of leiomyoma cystic degeneration and myxoid degeneration, the relation between T2 weighted signal intensity and diffusion is similar to that of a normal gallbladder, i.e., T2 weighted signal intensity highly hyperintense without diffusion restriction.

Table-1 supports our notion that, when a tissue has T2 relaxation time in the range of spleen T2, longer T2 'depresses' diffusion measure. However, when a tissue has a T2 relaxation time much longer than spleen T2, this depression effect of longer T2 is lost, or even longer T2 may be associated with higher diffusion measures.

For most of the tumors, much longer T2 may suggest a tumor is waterier. Inappropriate use of diffusion gradients (*b*-values) may exaggerate T2 relaxation's contribution to ADC calculation. Some examples of parotid tumors are as illustrated in Fig-1. Fig-1 is based on our parotid tumor literature search results where we searched for articles reported both quantitative ADC measure and T2 relaxation value or quantitative T2 signal intensity.

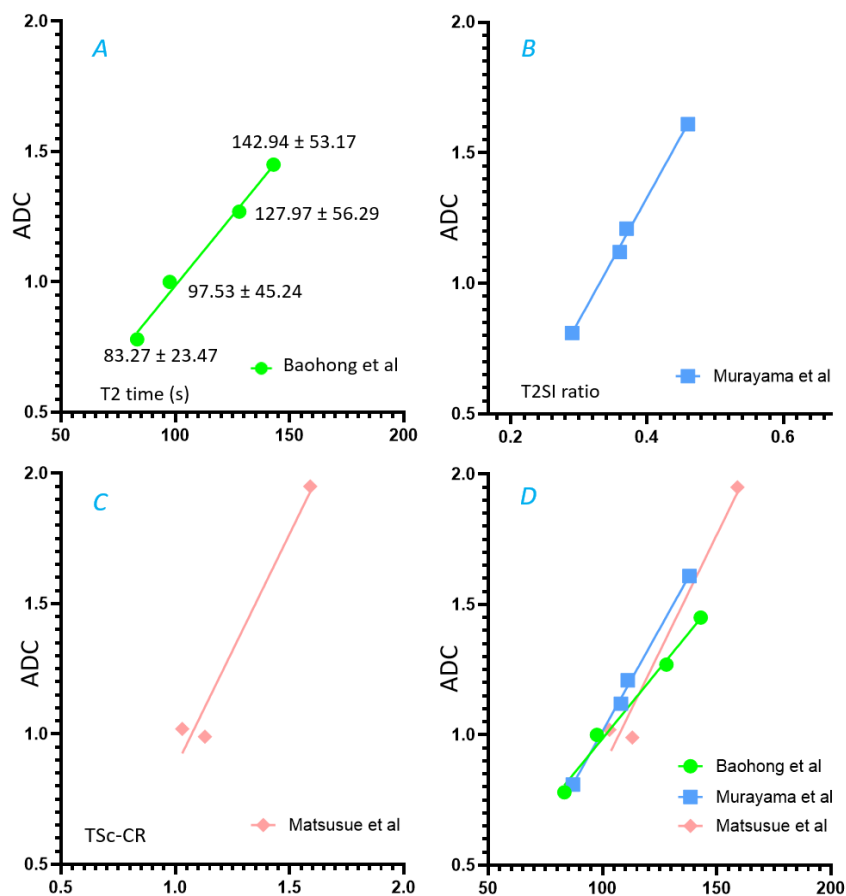

**Fig-1. Almost linear relationship between parotid tumor T2 relaxation time (or T2 weighted image signal) and parotid tumor ADC (×10⁻³ mm²/s).** A: Data from Baohong *et al*. [7], 3T scanner, ADC were from two b-values (0 and 1000 s/mm²). ADC from lower to higher ranking: Warthin's tumor, malignant tumor, benign tumor, pleomorphic adenoma. B: Data from Murayama *et al*. [8], 3T scanner, ADC were from three b-values (0, 500, 800 s/mm²). T2SI ratio: T2 weighted image signal ratio of tumor to the cerebrospinal fluid. ADC from lower to higher ranking: Warthin's tumor, parotid cancer, basal cell adenoma, pleomorphic adenoma. C: Data from Matsusue *et al*. [9], 1.5 Scanner, ADC were from two b-values (0 and 800 s/mm²). TSc-CR: tumor to spinal cord contrast ratio on T2 weighted image. ADC from lower to higher ranking: Warthin's tumor; malignant tumor, pleomorphic adenoma. D: aggregation of A, B, and C data, with Y-axes of B and C re-scaled to that of A.

Fig-1 shows almost perfect linear relationships between parotid tumor T2 relaxation time (or T2 weighted image signal) and parotid tumor ADC. This is unexpected. We propose a possible explanation (Fig-2, for the example of data of Baohong *et al*. [7]). On *b*=0 images, the signal of tumor (relative to other tissues) can be dominated by T2 effect. On *b*=1000 (or *b*=800) images, it is possible to this *b*-value setup is too high, and tumor regions on *b*=1000 (or *b*=800) images is dominated by noises, and thus there is not much difference among these four tumors (Fig-2). In the end, the ADCs, which are determined by the slope between the signal at *b*=0 and the signal at *b*=1000, are also dominated by the T2 effect. In fact, since there is no diffusion gradient with *b*=0 images, and there are only noises on *b*=1000 (or *b*=800), ADC map is devoid of diffusion information.

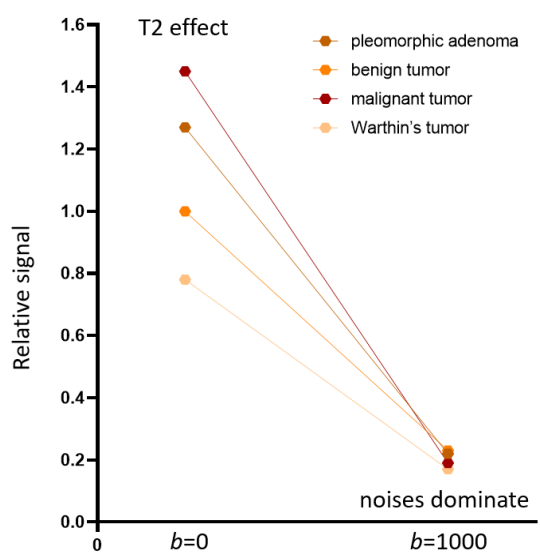

**Fig-2. If the parotid tumor signal on largest *b*-value image is dominated by noises, then T2 relaxation contribution to ADC calculation may be exaggerated, with longer T2 being associated with higher ADC measure.** This Fig is illustrated for the example of data of Baohong *et al*. [7].

The Fig-1 results agree with the pattern of scenarios [1,2] in table-1, while oppose the pattern of scenarios [3-12] in table-1. These phenomena may suggest that, when a T2 relaxation value moves away from spleen T2's 61 ms (at 3T) and myometrium T2's of 79 ms (at 3T) toward even higher values, a longer T2 relaxation time may be associated with a higher diffusion measure (i.e., higher T2 relaxation time value is related closer to the ADC behavior of gallbladder). On the other hand, inappropriate selection of *b*-value as illustrated in Fig-2, can exaggerate the T2 contribution to ADC calculation.

The discussion in the letter shows the complicated relationships between T2 relaxation time and diffusion measures. When a T2 value is in the range of those of spleen and uterine myometrium, longer T2 'depresses' diffusion measure. When a T2 is much longer, longer T2 relaxation time is associated with higher diffusion measures.

**Conflict-of-interest**: Nil.